\documentclass[12pt]{article}

\usepackage[utf8]{inputenc}
\usepackage{latexsym, graphicx} 
\usepackage{amsmath,amsfonts,amssymb}
\usepackage{cancel}
\usepackage{mathrsfs}
\usepackage{tensor}
\usepackage{xcolor}
\usepackage{cite}
\usepackage{caption,subcaption}
\usepackage[colorlinks=true,linkcolor=blue,citecolor=blue,urlcolor=blue,filecolor=black]{hyperref}

\renewenvironment{thebibliography}[1]{
	\begin{oldthebibliography}{#1}
		\setlength{\itemsep}{2pt}
		\setlength{\parskip}{2pt}
	}
	{
	\end{oldthebibliography}
}

\numberwithin{equation}{section} 

\usepackage{pict2e}
\makeatletter
\newcommand{\loplus}{\mathbin{\mathpalette\dog@lsemi{+}}}
\newcommand{\dog@lsemi}[2]{\dog@semi{#1}{#2}{270,90}}
\newcommand{\dog@semi}[3]{%
	\begingroup
	\sbox\z@{$\m@th#1#2$}%
	\setlength{\unitlength}{\dimexpr\ht\z@+\dp\z@\relax}%
	\makebox[\wd\z@]{\raisebox{-\dp\z@}{%
			\begin{picture}(1,1)
			\linethickness{\variable@rule{#1}}
			\roundcap
			\put(0.5,0.5){\makebox(0,0){\raisebox{\dp\z@}{$\m@th#1#2$}}}
			\put(0.5,0.5){\arc[#3]{0.5}}
			\end{picture}%
	}}%
	\endgroup
}
\newcommand{\variable@rule}[1]{%
	\fontdimen8  
	\ifx#1\displaystyle\textfont3\else
	\ifx#1\textstyle\textfont3\else
	\ifx#1\scriptstyle\scriptfont3\else
	\scriptscriptfont3\relax
	\fi\fi\fi
}
\makeatother

\newcommand*{\dd}{\mathop{}\!d}

\newcommand{\zbar}{\bar{z}}
\newcommand{\wbar}{\bar{w}}
\newcommand{\partialbar}{\bar{\partial}}

\newcommand{\cO}{O}
\newcommand{\cB}{\mathcal{B}}
\newcommand{\Bh}{\mathcal{B}_h}
\newcommand{\Om}{\{O\}}

\begin{document}
	
	\begin{titlepage}
		\thispagestyle{empty}
		
		\begin{flushright}
		\end{flushright}
		
		\vskip3cm
		
		\begin{center}  
			{\Large\textbf{Virasoro blocks and the reparametrization formalism}}
			
			\vskip1cm
			
			\centerline{Kevin Nguyen}
			
			\vskip1cm
			
			{\it{Department of Mathematics, King's College London,\\
					The Strand, London WC2R 2LS, UK}}\\
			\vskip 0.5cm
			{kevin.nguyen@kcl.ac.uk}
			
		\end{center}
		
		\vskip1cm
		
		\begin{abstract} 
			An effective theory designed to compute Virasoro identity blocks at large central charge, expressed in terms of the propagation of a reparametrization/shadow mode between bilocal vertices, was recently put forward. In this paper I provide the formal theoretical framework underlying this effective theory by reformulating it in terms of standard concepts~: conformal geometry, generating functionals and Feynman diagrams. A key ingredient to this formalism is the bilocal vertex operator, or reparametrized two-point function, which is shown to generate arbitrary stress tensor insertions into a two-point function of reference. I also suggest an extension of the formalism designed to compute generic Virasoro blocks.
		\end{abstract}
		
	\end{titlepage}
	
	{\hypersetup{linkcolor=black}
		\tableofcontents
	}
	
	\section{Introduction}
	Recently an effective theory designed to efficiently compute Virasoro identity blocks at large central charge, also known as \textit{reparametrization formalism}, has been put forward in the context of two-dimensional conformal field theory (CFT) \cite{Cotler:2018zff,Haehl:2018izb,Haehl:2019eae,Anous:2020vtw,Nguyen:2021jja}. But even though it deals with the exchange of the identity operator and its Virasoro descendants, namely a sector of the theory fully controlled by conformal symmetry, the reparametrization formalism thus far mostly amounts to a series of prescriptions of rather obscure origins. The purpose of the present paper is to explain the foundations of this effective theory in terms of more standard concepts~: conformal geometry, generating functionals and Feynman diagrams. Some steps in that direction were already taken in a previous publication \cite{Nguyen:2021jja}. The fundamental understanding developed here also suggests a generalization of the reparametrization formalism allowing for the computation of more generic Virasoro blocks. Note that a different although somewhat related approach has been discussed in \cite{Vos:2021erh}. 
	
	One important motivation for the reparametrization formalism originates from the study of quantum chaos in the context of the AdS$_3$/CFT$_2$ holographic correspondence \cite{Shenker:2013pqa,Shenker:2014cwa,Roberts:2014ifa,Maldacena:2015waa}. The prototypical probe of quantum chaos is an out-of-time ordered four-point correlator, which can be obtained from a euclidean four-point function by analytic continuation. In particular it has been understood that holographic CFTs are maximally chaotic which in turns implies that the relevant four-point function is dominated by the Virasoro identity block in the limit of large central charge. My own perspective is that the reparametrization formalism is as close as it gets to a dual gravitational description, and indeed some of its basic ingredients have appeared naturally from gravitational considerations \cite{Cotler:2018zff,Fitzpatrick:2016mtp,DHoker:2019clx,Nguyen:2021pdz}. The reason for this is further elucidated in this work. The basic ingredients of the reparametrization formalism, namely the bilocal vertex operators as well as the effective action for the `reparametrization mode' (also known as `shadow mode' or `scramblon' depending on the context), will be shown to be generating functionals for correlators of the form $\langle OOT\,...\,T\rangle$ and $\langle T\,...\,T \rangle$ with an arbitrary number of stress tensor insertions. Since the holographic dictionary precisely relies on the identification of generating functionals on both sides of the duality \cite{Gubser:1998bc,Witten:1998qj}, the latter indeed constitute the preferred language. 
	
	In order to set the stage, let me briefly review the rules prescribed with the reparametrization formalism \cite{Haehl:2019eae,Anous:2020vtw}. For concreteness I focus here on the four-point function involving two pairs of identical primary operators $V$ and $W$.\footnote{The discussion straightforwardly generalizes to \textit{star channel} contribution \cite{Anous:2020vtw} to correlators $\langle \prod_{i=1}^n O_i O_i \rangle$ of $n$ pairs of identical operators with ($O_i\neq O_j$ for $i\neq j$). On the other hand contributions from \textit{comb channels} are obtained by gluing of lower-point Virasoro identity blocks.} As is well-known the latter can be decomposed into a sum of Virasoro blocks $\mathcal{V}_h(u)$,
	\begin{equation}
	\label{eq:Virasoro blocks}
	\langle V(1) V(2) W(3) W(4) \rangle=\frac{1}{(z_{12})^{2h_V}(z_{34})^{2h_W}} \sum_{O} C_{VVO}\, C_{WWO}\, \mathcal{V}_{h_O}(u)\,,
	\end{equation}
	where $u=\frac{z_{12}z_{34}}{z_{13}z_{24}}$ and the sum runs over all primary operators $O$ in the spectrum. Each term in the sum describes the exchange of a primary operator and its Virasoro descendants between the two pairs of external operators, with $C_{VVO}$ and $C_{WWO}$ the corresponding three-point fusion coefficients. The reparametrization formalism allows to compute the Virasoro identity block $\mathcal{V}_0$ in the following way. First one introduces the bilocal vertex operator
	\begin{equation}
	\label{bilocal}
	\mathcal{B}_{h_V}(1,2)\equiv \left( \frac{\partial f(z_1,\zbar_1)\, \partial f(z_2,\zbar_2)}{\left(f(z_1,\zbar_1)-f(z_2,\zbar_2)\right)^2}\right)^{h_V}\,, 
	\end{equation}
	which coincides with the expression of the two-point function $\langle V(1) V(2) \rangle$ after a conformal transformation $z \mapsto f(z)$ from the plane, with the difference that $f$ is not necessarily holomorphic. Second one expands this function in terms of the `reparametrization mode' $\epsilon(z,\zbar)$ via 
	\begin{equation}
	\label{eq:f intro}
	f=e^{\epsilon\, \partial_z} z=z+\epsilon+\frac{1}{2} \epsilon \partial \epsilon + O(\epsilon^3)\,.
	\end{equation}
	In this approach the reparametrization mode is considered the appropriate dynamical fluctuation for a perturbative expansion in the limit of large central charge. Its dynamics is modeled over that of the stress tensor through the differential relation
	\begin{equation}
	\label{eq:epsilon propagators}
	\langle \partial^3 \epsilon(z_1)\, ...\, \partial^3 \epsilon(z_n) \rangle = \left(\frac{12}{c}\right)^n\langle T(z_1)\, ... \, T(z_n) \rangle\,.
	\end{equation}
	The stress tensor correlators are fully determined from the conformal Ward identity \cite{Belavin:1984vu}, and one needs to invert the derivative operators on the left-hand side of \eqref{eq:epsilon propagators} in order to determine the `auxiliary' correlators $\langle \epsilon(z_1)\, ...\, \epsilon(z_n) \rangle$. This can be done using formula \eqref{inverse operator} below although this yields tedious iterated integrals. In practice it is simpler to look for distributions $\langle \epsilon(z_1)\, ...\, \epsilon(z_n) \rangle$ which have the right singularity structure in the form of poles and branch cuts, such as to satisfy \eqref{eq:epsilon propagators} upon derivation. Away from these singularities the correlators are holomorphic. The simplest example is that of the two-point function \cite{Haehl:2019eae,Anous:2020vtw,Nguyen:2021jja}
	\begin{equation}
	\label{epsilon epsilon}
	\langle \epsilon(z_1)\epsilon(z_2) \rangle =\frac{6}{c}\, (z_{12})^2 \ln z_{12}\,, 
	\end{equation}
	while the three-point function has been worked out in \cite{Anous:2020vtw}.
	
	With these definitions at hand, the Virasoro identity block contribution to \eqref{eq:Virasoro blocks} is conjectured to be equal to
	\begin{equation}
	\label{identity block}
	\mathcal{V}_0(u)  =(z_{12})^{2h_V}(z_{34})^{2h_W}\langle \cB_{h_V}(1,2) \cB_{h_W}(3,4) \rangle_c \,.
	\end{equation}
	The practical meaning of the right-hand side of \eqref{identity block} is the following~: expand the bilocal vertices in powers of the reparametrization mode, and use the correlators  \eqref{eq:epsilon propagators} to evaluate each term in the resulting sum, keeping only connected contributions. The fact that \eqref{eq:epsilon propagators} features a power of $1/c$ naturally yields a perturbative structure in the limit of large central charge \cite{Cotler:2018zff,Haehl:2018izb,Haehl:2019eae,Anous:2020vtw,Nguyen:2021jja}.
	
	In this paper I will revisit the reparametrization formalism reviewed above using first principles. I start in section~\ref{sec:generating functionals} by showing that the bilocal vertex operator \eqref{bilocal} is in fact the generating functional for the correlators $\langle VV T\,...\, T\rangle$ with an arbitrary number of stress tensor insertions while the function $f(z,\zbar)$, known as quasi-conformal mapping, parametrizes deformations of the underlying conformal geometry. In doing so the validity of the formula $\eqref{eq:f intro}$ will be questioned, while a potentially more appropriate replacement will be put forward. These results are then used in section~\ref{sec:identity blocks} to show that the prescription \eqref{eq:epsilon propagators}-\eqref{identity block} amounts to a textbook computation of Feynman diagrams with an arbitrary number of stress tensor `particles' exchanged between the external operators, thereby completing the argument given in \cite{Nguyen:2021jja}. Finally in section~\ref{sec:generic blocks} I suggest a generalization of the reparametrization formalism which would allow the computation of more generic Virasoro blocks $\mathcal{V}_h$, although its practicality is still to be determined.  
	
	\paragraph{Conventions.} We work in euclidean signature. We use the shorthand notations $T\equiv -2\pi T_{zz}$ for the holomorphic component of the stress tensor, $\delta(z)\equiv \delta^{(2)}(z,\zbar)$ for the delta distribution normalized as $\int d^2z\, \delta(z)=1$ with $d^2z=1/2 \dd z \dd \zbar$, and $z_{ij}\equiv z_i-z_j$ for relative distances. For ease of notation we sometimes suppress coordinate dependence, and write expressions such as $\partialbar \epsilon_1 \equiv \partial_{\zbar_1} \epsilon(z_1,\zbar_1)$. We make repetitive use of the magic distributional identity
	\begin{equation}
	\label{eq:magic}
	\partial_{\zbar}\left( \frac{1}{z}\right)=\pi \delta(z)\,,
	\end{equation}
	and similarly use the inverse operator
	\begin{equation}
	\label{inverse operator}
	\partial_{\zbar}^{-1} g(z,\zbar)= \frac{1}{\pi} \int d^2w\, \frac{1}{z-w}\, g(w,\wbar)\,.
	\end{equation}
	
	\section{Generating functionals for stress tensor insertions}
	\label{sec:generating functionals}
	We start by considering the generating functional that allows to compute the effect of an arbitrary number of stress tensor insertions into a given correlation function of reference. It is formally given by
	\begin{equation}
	\label{def:generating functional}
	Z\left[\mu \right] = \sum_{n=0}^\infty \frac{(-\pi)^{-n}}{n!} \int d^2w_1...\,d^2w_n\, \mu(w_1,\wbar_1)...\mu(w_n,\wbar_n) \langle \Om \, T(w_1)...T(w_n)  \rangle\,,
	\end{equation}
	where $\Om$ is a short-hand notation for a string of $m$ primary operators of reference,
	\begin{equation}
	\label{O string}
	\Om \equiv \cO_1(z_1)...\cO_m (z_m)\,.
	\end{equation}
	The background field $\mu$ in \eqref{def:generating functional} is the source conjugated to the stress tensor. One can differentiate \eqref{def:generating functional} with respect to $\mu$ in order to generate correlation functions with an arbitrary number of stress tensor insertions,
	\begin{equation}
	\label{eq:funct diff}
	\langle \Om\, T(w_1) \, ...\, T(w_n) \rangle = (-\pi)^n\, \frac{\delta^n Z}{\delta \mu(w_1,\wbar_1)\, ...\, \delta \mu(w_n,\wbar_n)}\bigg|_{\mu=0}\,. 
	\end{equation}
	Of course the source conjugated to the stress tensor is the background conformal geometry itself, with underlying metric
	\begin{equation}
	\label{eq:deformed metric}
	ds^2=dz\, d\zbar+ \mu(z,\zbar)\, d\zbar^2\,.
	\end{equation}
	This form of the metric is sometimes called chiral or lightcone gauge \cite{Polyakov:1987zb,Polyakov:1988qz}.
	It is furthermore very useful to parametrize the metric deformation $\mu(z,\zbar)$ in terms of a function $f(z,\zbar)$ defined implicitly as the solution of the Beltrami equation
	\begin{equation}
	\label{eq:Beltrami}
	\mu=\frac{\partialbar f}{\partial f}\,.
	\end{equation}
	Written in this form, the source $\mu(z,\zbar)$ is a \textit{Beltrami differential} constructed out of the \textit{quasi-conformal mapping} $f(z,\zbar)$ \cite{Lehto1986UnivalentFA,Donaldson} (see also the physics reviews \cite{Alvarez:1985ez,Nelson:1986ab,Giddings:1987im,DHoker:1988pdl}). For a quasi-conformal mapping infinitesimally close to the identity, we can parametrize the solution of the Beltrami equation \eqref{eq:Beltrami} in terms of a vector field $\epsilon(z,\zbar)$ \cite{DHoker:1988pdl},
	\begin{equation}
	\label{eq:epsilon}
	\mu=\partialbar \epsilon\,.
	\end{equation}
	The Beltrami equation \eqref{eq:Beltrami} can then be solved as a perturbative series \cite{Lehto1986UnivalentFA}
	\begin{equation}
	\label{Neumann}
	f(z,\zbar)=z+ \sum_{i=0}^\infty \partialbar^{-1}\varphi_i(z,\zbar)\,, 
	\end{equation}
	where the functions $\varphi_i(z,\zbar)$ satisfy the recursive relations
	\begin{equation}
	\varphi_0=\mu\,, \qquad \varphi_i=\mu\, \partial \partialbar^{-1} \varphi_{i-1}\,, \quad i>0\,.
	\end{equation}
	Inserting \eqref{eq:epsilon} into \eqref{Neumann}, we thus have
	\begin{equation}
	\label{eq:f expansion}
	f(z,\zbar)=z+\epsilon+\partialbar^{-1}\left( \partialbar \epsilon\, \partial \epsilon\right)+O(\epsilon^3)\,.
	\end{equation}
	This equation is analogous to \eqref{eq:f intro} in that $\epsilon$ appears as the linear perturbation around the identity map $z \mapsto f(z,\zbar)=z$, however the higher order completion differs. From \eqref{eq:epsilon} we see that functional differentiation with respect to the `reparametrization mode' $\epsilon$ alternatively produces insertions of $\partialbar T$,
	\begin{equation}
	\label{eq:funct diff dT}
	\langle \Om\, \partialbar T(w_1) \, ...\, \partialbar T(w_n) \rangle = \pi^n\, \frac{\delta^n Z}{\delta \epsilon(w_1,\wbar_1)\, ...\, \delta \epsilon(w_n,\wbar_n)}\bigg|_{\epsilon=0}\,. 
	\end{equation}
	The corresponding correlation functions with $T$ in place of $\partialbar T$ can be obtained by acting with the inverse operator $\partialbar^{-1}$ given in \eqref{inverse operator}.
	
	As is well-known, Virasoro symmetry fully controls stress tensor insertions through the conformal Ward identity \cite{Belavin:1984vu}. As explicitly demonstrated in the appendix, this in turn translates into a Ward identity for the generating functional $Z[\mu]$. The latter can be decomposed into the product 
	\begin{equation}
	Z[\mu]=Z_0[\mu]\, Z_c[\mu]\,,
	\end{equation}
	where $Z_0[\mu]$ generates the vacuum correlators $\langle T\, ...\, T \rangle$ while $Z_c[\mu]$ generates the \textit{connected} correlators $\langle \Om\, T\, ...\, T \rangle_c$. They separately satisfy the Ward identities
	\begin{align}
	\label{Psi0 Ward id}
	\partial f(z,\zbar)\, \frac{\delta Z_0[\mu]}{\delta f(z,\zbar)}&=\frac{c}{12\pi}\, \partial^3 \mu(z,\zbar)\, Z_0[\mu]\,,\\
	\label{Psic Ward id}
	\partial f(z,\zbar)\, \frac{\delta Z_c[\mu]}{\delta f(z,\zbar)}&=-\sum_{i=1}^m \left(h_i\, \partial_z\delta(z-z_i)-\delta(z-z_i)\partial_{z_i}\right) Z_c[\mu]\,.
	\end{align}
	I now discuss the solutions to these equations. Details of computations can be found in the appendix.
	
	\paragraph{Vacuum generating functional.} The generating functional $Z_0[\mu]$ solution to \eqref{Psi0 Ward id} is explicitly given by
	\begin{equation}
	\label{eq:Psi no puncture}
	Z_0\left[\mu\right]=\exp\left[-\frac{c}{24\pi} \int d^2z\,  \frac{\partialbar f}{\partial f}\, \partial^2 \ln \partial f\right]\equiv e^{-W_0[\mu]}\,,
	\end{equation}
	where the relation between $f$ and $\mu$ is implicitly given through the Beltrami equation \eqref{eq:Beltrami}. This expression has appeared in various forms in the literature \cite{Alekseev:1988ce,Yoshida:1988xm,Verlinde:1989ua,Lazzarini:1990xid,Aldrovandi:1996sa}. See \cite{Nguyen:2021pdz} for a detailed discussion. To explicitly compute stress tensor correlations from $Z_0[\mu]$, it is best to work in terms of the `reparametrization mode' $\epsilon$, substituting the series expansion \eqref{eq:f expansion} into \eqref{eq:Psi no puncture} and applying the formula \eqref{eq:funct diff dT}. In practice one needs to truncate this expansion at an appropriate order that allows to compute the correlators of interest.
	
	As an illustration but also in order to correct the naive treatment presented in \cite{Nguyen:2021jja}, let us derive the one-, two- and three-point stress tensor correlators. Since these do not contain disconnected contributions, we can as well use $W_0[\mu]$ to generate them more directly. The expansion of the latter to order $O(\epsilon^3)$ is
	\begin{equation}
	\label{eq:cubic action}
	W_0=\frac{c}{24\pi} \int d^2z \left(\partialbar \epsilon\, \partial^3 \epsilon-\partialbar \epsilon\, \partial^2 \epsilon\, \partial^2 \epsilon +O(\epsilon^4)\right),
	\end{equation}
	where heavy use of integration by parts has been made. From this, one obtains the standard expressions for the correlators on the plane,\footnote{One simple way to proceed is to use the magic formula \eqref{eq:magic} in order to express \eqref{eq:cubic action} as a nonlocal functional of the field $\mu=\partialbar \epsilon$ alone. For the quadratic part of the generating functional for instance, we have
		\begin{equation}
		\nonumber
		\int d^2z_1\, \partial^3 \epsilon_1\, \partialbar \epsilon_1=\frac{1}{\pi} \int d^2z_1 d^2z_2\, \partialbar_1 \left(\frac{1}{z_{12}}\right)  \partial^3 \epsilon_1\, \partialbar \epsilon_2=\frac{1}{\pi} \int d^2z_1 d^2z_2\, \partial_1^3 \left(\frac{1}{z_{12}}\right)  \partialbar \epsilon_1\, \partialbar \epsilon_2\,.
		\end{equation}}
	\begin{equation}
	\begin{split}
	\langle T(z) \rangle&=\pi\, \frac{\delta W_0}{\delta \mu}\Big|_{\mu=0}=0\,,\\
	\label{TT}
	\langle T(z_1) T(z_2) \rangle&=-\pi^2\, \frac{\delta^2 W_0}{\delta \mu_1\, \delta \mu_2}\Big|_{\mu=0}= \frac{c}{2z_{12}^4}\,,\\
	\langle T(z_1) T(z_2) T(z_3) \rangle&=\pi^3\, \frac{\delta^3 W_0}{\delta \mu_1\, \delta \mu_2\, \delta \mu_3}\Big|_{\mu=0}= \frac{c}{z_{12}^2z_{13}^2z_{23}^2}\,,
	\end{split}
	\end{equation}
	where it is understood that $\mu=\partialbar \epsilon$ and the condition $\mu(z,\zbar)=0$ corresponds to $\epsilon(z,\zbar)=0$. An equivalent derivation is given in \cite{Lazzarini:1990xid}. Let us stress that the expansions \eqref{eq:f intro} and \eqref{eq:f expansion} for $f$ cannot be used interchangeably. Indeed plugging the expansion \eqref{eq:f intro} into \eqref{eq:Psi no puncture} would instead yield
	\begin{equation}
	W_0=\frac{c}{24\pi} \int d^2z \left(\partialbar \epsilon\, \partial^3 \epsilon+\frac{1}{2}\partialbar \epsilon\, \partial^2 \epsilon\, \partial^2 \epsilon +O(\epsilon^4)\right)\,,
	\end{equation}
	which, although very similar to \eqref{eq:cubic action}, differs by a coefficient and is therefore incorrect.
	
	Note that the quadratic term in the effective action \eqref{eq:cubic action} has been considered in the reparametrization formalism as a way to determine the propagator of the reparametrization mode $\epsilon$ \cite{Cotler:2018zff,Haehl:2018izb,Haehl:2019eae}. In the presentation given here, the latter is not a dynamical entity but rather plays the role of a background source for the stress tensor. Of course the two viewpoints are related by a Legendre transform (see for example chapter 11.3 of \cite{Peskin:1995ev}). 
	
	\paragraph{Bilocal generating functional.}
	We now consider the generating functional $Z_{2,c}[\mu]$ for the connected correlation functions $\langle \cO \cO T\, ... \, T \rangle_c$ with a primary operator $O$ of conformal weight $h$. It is given by the solution of the Ward identity \eqref{Psic Ward id}, namely
	\begin{equation}
	\label{Psi2}
	Z_{2,c}[\mu]=\left(\frac{\partial f(z_1,\zbar_1)\, \partial f(z_2,\zbar_2)}{\left(f(z_1,\zbar_1)-f(z_2,\zbar_2) \right)^2} \right)^h\equiv \Bh(1,2)\,.
	\end{equation}
	We have recovered in this way the \textit{bilocal vertex operator} or reparametrized two-point function \eqref{bilocal}. Let us explicitly verify that correlation functions with various stress tensor insertions are correctly reproduced by functional differentiation of \eqref{Psi2}. To this end we expand it perturbatively around $f(z,\zbar)=z$ using \eqref{eq:f expansion},
	\begin{align}
	\label{bilocal expansion}
	\Bh(1,2)=\frac{1}{(z_{12})^{2h}} \sum_n \Bh^{(n)}(1,2)\,,
	\end{align}
	where the first few terms are given by
	\begin{equation}
	\begin{split}
	\Bh^{(0)}(1,2)&=1\,,\\
	\Bh^{(1)}(1,2)&=b^{(1)}_h(1,2)\,,\\
	\Bh^{(2)}(1,2)&=\frac{1}{2!}\left(b^{(1)}_h(1,2)\right)^2+b^{(2)}_h(1,2)+R_h^{(2)}(1,2)\,,\\
	&\hskip 2mm \vdots\\
	\Bh^{(n)}(1,2)&=\frac{1}{n!}\left(b^{(1)}_h(1,2)\right)^n + \text{lower orders in } h\,,
	\end{split}
	\end{equation}
	with
	\begin{equation}
	\label{eq:b vertices}
	\begin{split}
	b^{(1)}_h(1,2)&=h\left(\partial \epsilon_1+\partial \epsilon_2-\frac{2(\epsilon_1-\epsilon_2)}{z_{12}}\right)\,,\\
	b^{(2)}_h(1,2)&=h\left(\frac{\epsilon_1\partial^2 \epsilon_1+\epsilon_2\partial^2 \epsilon_2}{2}-\frac{\epsilon_1\partial \epsilon_1-\epsilon_2\partial \epsilon_2}{z_{12}}+\frac{(\epsilon_1-\epsilon_2)^2}{z_{12}^2}\right)\,,\\
	R^{(2)}_h(1,2)
	&=h \left(\left(\frac{\partial_1}{2}- \frac{1}{z_{12}}\right)\partialbar_1^{-1}(\partial \epsilon_1 \partialbar \epsilon_1-\epsilon_1 \partial\partialbar \epsilon_1)+(z_1 \leftrightarrow z_2) \right)\,.
	\end{split}
	\end{equation}
	Using the above expansion, equation \eqref{eq:funct diff dT} yields correlation functions with insertions of~$\partialbar T$. For example we find
	\begin{equation}
	\label{OO dT}
	\langle \cO(z_1) \cO(z_2) \partialbar T(w) \rangle=-\frac{\pi }{(z_{12})^{2h}}\left[h\left(\partial_w +\frac{2}{z_{12}}\right) \delta(w-z_1)+(z_1 \leftrightarrow z_2)\right]\,,
	\end{equation}
	in agreement with the diffeomorphism Ward identity \eqref{eq:diff Ward}.
	
	Note that even though we have used the expansion \eqref{eq:f expansion} rather than \eqref{eq:f intro}, the expansion \eqref{bilocal expansion} agrees with the expression in \cite{Anous:2020vtw} up to an additional remainder term $R_h^{(2)}$. We can expect the latter to be negligible in the computation of Virasoro blocks as no such term was needed in \cite{Nguyen:2021jja} in order to find agreement with a more traditional approach based on Feynman diagrams. Indeed the rules \eqref{eq:epsilon propagators}-\eqref{identity block} imply that $R^{(2)}_h$ produces terms like
	\begin{equation}
	\partialbar_1^{-1}\left(\partial_1 \langle \epsilon_1\,...\, \rangle\, \partialbar_1 \langle \epsilon_1\,...\, \rangle \right)\,,
	\end{equation}
	which vanish since $\epsilon$-correlators are holomorphic away from a set of singularities as discussed in the introduction.\footnote{Instances where $\epsilon$-correlators need to be evaluated at these singularities are precisely those where UV divergences occur in the reparametrization formalism \cite{Nguyen:2021jja}. Their treatment is an open problem which goes beyond the present work. Further comments are given at the end of section~\ref{sec:identity blocks}.} 
	
	It is also worth mentioning that the expression \eqref{Psi2} with $f(z,\zbar)=f(z)$ a holomorphic function has appeared as the gravitational Wilson line in AdS$_3$ evaluated in a state with stress tensor vacuum expectation value equal to the Schwarzian dervative $\langle T(z) \rangle=\frac{c}{12} S[f\,;z]$ \cite{Fitzpatrick:2016mtp,DHoker:2019clx}. Such a state is obtained from the vacuum by conformal transformation $z \mapsto f(z)$. It is very likely that \eqref{Psi2} is the expression for this same Wilson line when an arbitrary boundary source $\mu$ is turned on, just like $W_0[\mu]$ is the gravitational onshell action with arbitrary boundary source \cite{Nguyen:2021pdz}. 
	
	\paragraph{Trilocal generating functional.} In preparation for a generalization of the reparametrization formalism to be discussed in section~\ref{sec:generic blocks}, let us also consider the generating functional for the connected correlators $\langle \cO_1 \cO_2 \cO_3\, T\, ... \, T \rangle_c$. It is given by the reparametrized three-point function,
	\begin{equation}
	\label{Psi3}
	Z_{3,c}[\mu]=C_{123}\,\frac{(\partial f_1)^{h_1} (\partial f_2)^{h_2} (\partial f_3)^{h_3}}{(f_1-f_2)^{h_1+h_2-h_3}(f_1-f_3)^{h_1+h_3-h_2}(f_2-f_3)^{h_2+h_3-h_1}}\,.
	\end{equation}
	As before we can expand it perturbatively around $f(z,\zbar)=z$. We find
	\begin{equation}
	Z_{3,c}=\langle \cO_1 \cO_2 \cO_3 \rangle\, \sum_n Z_{3,c}^{(n)}\,,
	\end{equation}
	where the reference three-point function is
	\begin{equation}
	\langle \cO_1 \cO_2 \cO_3 \rangle=\frac{C_{123}}{(z_{12})^{h_1+h_2-h_3}(z_{13})^{h_1+h_3-h_2}(z_{23})^{h_2+h_3-h_1}}\,,
	\end{equation}
	while the first two terms of the expansion are given by
	\begin{equation}
	\begin{split}
	Z_{3,c}^{(0)}&=1\,,\\
	Z_{3,c}^{(1)}&=\left[h_1 \partial \epsilon_1- \left(\frac{h_1+h_2-h_3}{z_{12}}+\frac{h_1+h_3-h_2}{z_{13}}\right)\epsilon_1\right]+(231)+(312)\,,
	\end{split}
	\end{equation}
	and $(213)$ and $(312)$ refer to cyclic permutations of the indices. Again we can apply \eqref{eq:funct diff dT} to explicitly compute the effect of inserting $\partialbar T$ into the correlator $\langle \cO_1 \cO_2 \cO_3 \rangle$, and we 
	find agreement with the diffeomorphism Ward identity \eqref{eq:diff Ward}.
	
	\section{Virasoro identity blocks}
	\label{sec:identity blocks}
	Having understood that the bilocal vertex $Z_{2,c}\equiv \Bh$ of the reparametrization formalism is actually the generating functional of the correlators $\langle \cO \cO\, T\,...\, T \rangle_c$, we are now in a position to understand the prescriptions reviewed in the introduction as evaluating the set of connected Feynman diagrams with stress tensor exchanges between the external operators $V,W$. Evidence for this was already given in a previous paper \cite{Nguyen:2021jja}.
	
\begin{figure}
		\centering
		\includegraphics[scale=0.5]{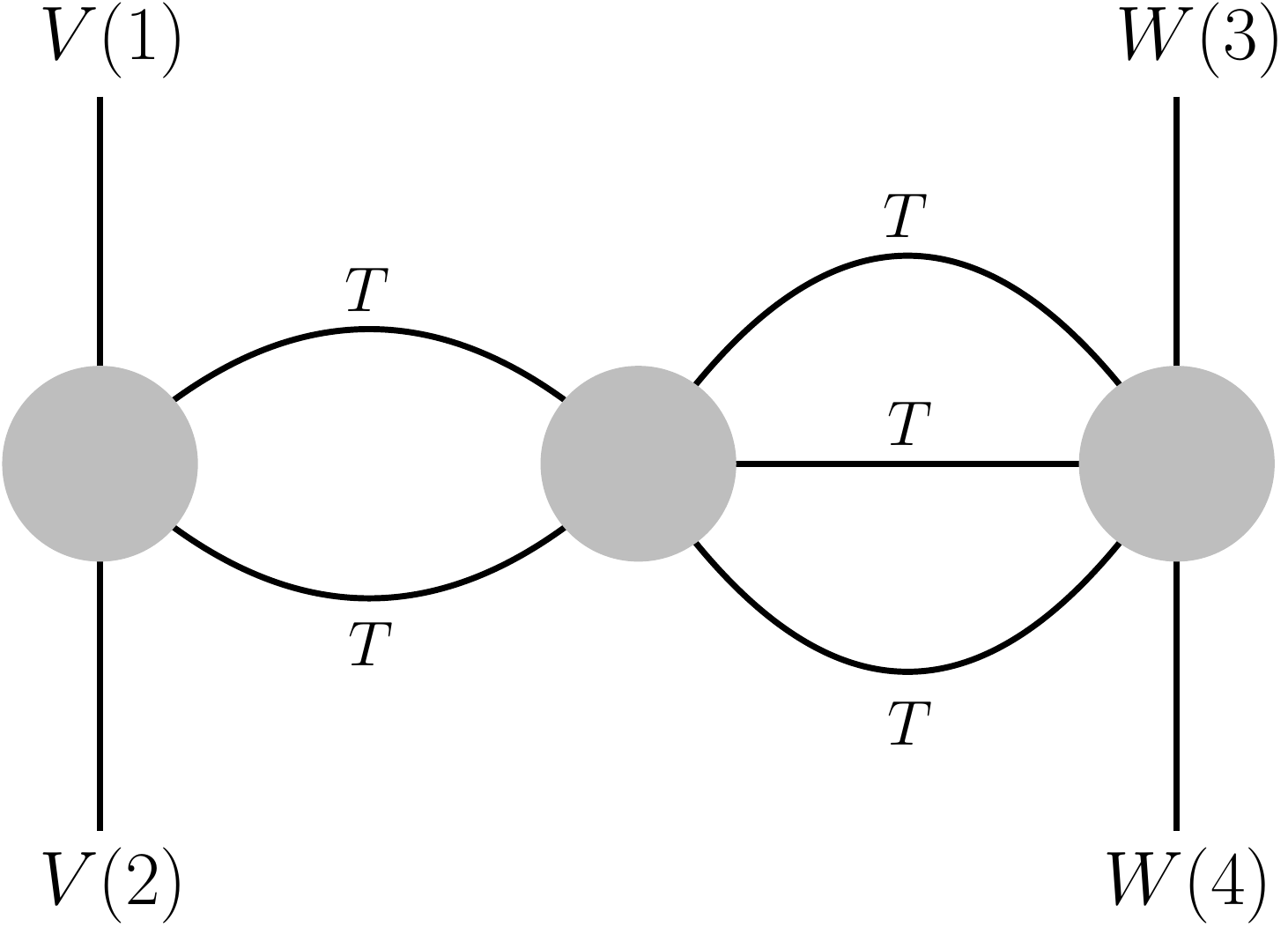}
		\caption{Feynman diagrams corresponding to stress tensor exchanges between two pairs of identical operators. The large grey circles  refer to \textit{exact} vertices $\langle VV \hat{T}... \hat{T}\rangle$ and $\langle \hat T\,...\, \hat T\rangle$.}
		\label{fig:identity block}
	\end{figure}

	Thus we aim at evaluating the Virasoro identity block \eqref{identity block}
	by expanding $\Bh$ in powers of the reparametrization mode $\epsilon$ and inserting its correlators \eqref{eq:epsilon propagators}. As previously mentioned, in the picture developed here $\epsilon$ plays the role of a source rather than that of a dynamical field, and the meaning of the quantities $\langle \epsilon\,...\,\epsilon \rangle$ is really that of stress tensor correlators trough the defining relation \eqref{eq:epsilon propagators}. We shall start by considering the closely related quantity
	\begin{align}
	\label{tilde V0}
	\tilde{\mathcal{V}}_0=\sum_{n,m} \frac{1}{n!m!}\int dM\, \delta^{(n)} \mathcal{B}_{h_V}(1,2)\, \langle \epsilon(z_1)\,...\,\epsilon(z_n)\epsilon(w_1)\,...\, \epsilon(w_m)\rangle_{\tilde c}\, \delta^{(m)} \mathcal{B}_{h_W}(3,4)\,,
	\end{align}
	with the short-hand notation
	\begin{equation}
	\delta^{(n)} \mathcal{B}_{h_V}(1,2)\equiv \frac{\delta^n \mathcal{B}_{h_V}(1,2)}{\delta \epsilon(z_1)\,...\,\delta \epsilon(z_n)}\bigg|_{\epsilon=0}\,,
	\end{equation}
	and integral measure given by
	\begin{equation}
	dM=\prod_{i=1}^n  d^2z_i \prod_{j=1}^m d^2w_j\,.
	\end{equation}
	The connectivity condition $(\tilde c)$ only discards disconnected correlators that attach to the \textit{same} bilocal vertex, in such a way that $\tilde{\mathcal{V}}_0$ is indeed overall connected. Said differently, at least one connected factor of the $\epsilon$-correlator must carry $z$ \textit{and} $w$ dependences, thereby ensuring that the two bilocal vertices are indeed connected to one another. For example in $\langle \epsilon(z_1)\epsilon(z_2) \epsilon(w_1) \epsilon(w_2)\rangle_{\tilde c}$ we discard
	$ \langle \epsilon(z_1)\epsilon(z_2) \rangle\, \langle \epsilon(w_1) \epsilon(w_2) \rangle $ while 
	$
	\langle \epsilon(z_1)\epsilon(w_1) \rangle\, \langle \epsilon(z_2) \epsilon(w_2) \rangle 
	$
	is retained. Plugging in \eqref{eq:funct diff dT}, \eqref{tilde V0} thus becomes
	\begin{align}
	\label{tilde V0 2}
	\tilde{\mathcal{V}}_0=\sum_{n,m} \frac{1}{n!m!}\int dM\, \langle VV \partialbar T(z_1)\,... \,\rangle_c\, \langle \epsilon(z_1)\,...\,\epsilon(w_m)\rangle_{\tilde c}\, \langle \,...\, \partialbar T(w_m) WW \rangle_c\,.
	\end{align}
	Using \eqref{eq:epsilon propagators} and integration by parts, we obtain the alternative expression
	\begin{align}
	\label{Feynman sum}
	\tilde{\mathcal{V}}_0=\sum_{n,m} \frac{1}{n!m!}\int dM\, \langle VV \hat T(z_1)... \hat T(z_n)\rangle_c\, \langle T(z_1)...T(w_m)\rangle_{\tilde c}\, \langle \hat T(w_1)... \hat T(w_m) WW \rangle_c\,,
	\end{align}
	where the hatted stress tensor is defined as
	\begin{equation}
	\label{hat T}
	\hat T(z)\equiv -\frac{12}{\pi c}\, \partial^{-3} \partialbar T(z)\,.
	\end{equation}
	Note that the latter can also be interpreted as the shadow of the stress tensor \cite{Haehl:2019eae,Banerjee:2022wht}.
	Here we see appear the \textit{partially amputated} correlators $\langle VV \hat T\, ...\, \hat T \rangle$ \cite{Nguyen:2021jja}. Indeed we can verify
	\begin{equation}
	\langle\, ...\, T(z)\,...\, \rangle = \int d^2w\, \langle T(z) T(w) \rangle\, \langle \,...\, \hat T(w)\,...\, \rangle\,, 
	\end{equation}
	as follows from the magic identity \eqref{eq:magic} and the expression for the stress tensor two-point function \eqref{TT}. Looking once more at the expression \eqref{Feynman sum}, we now see that $\tilde{\mathcal{V}}_0$ is a textbook sum of Feynman diagrams in position space. The diagrams featured are those with an arbitrary number of stress tensor `particles' propagating between the external states $V,W$. See figure~\ref{fig:identity block}. A particularity is that the vertices $\langle VV \hat T\,...\, \hat T\rangle$ and propagators $\langle T\,...\,T\rangle$ are actually exact (as opposed to free).
	
	In order to complete the argument, we are left to show that the following equality holds,
	\begin{equation}
	\label{V0 equal V0}
	\mathcal{V}_0=(z_{12})^{2h_V}(z_{34})^{2h_W}\, \tilde{\mathcal{V}}_0\,,
	\end{equation}
	where $\mathcal{V}_0$ and $\tilde{\mathcal{V}}_0$ are given by \eqref{identity block} and \eqref{tilde V0}-\eqref{tilde V0 2}, respectively. The two quantities are obviously closely related as they both involve expanding the bilocal vertex $\Bh$ in powers of the reparametrization mode $\epsilon$ as well as using the correlators \eqref{eq:epsilon propagators} of the latter. The gist of the proof of \eqref{V0 equal V0} is the following. Consider the correlators $\langle VV \partialbar T\,...\, \partialbar T \rangle$ appearing in \eqref{tilde V0 2}. They contain precisely one delta function per integration variable, acted on by some differential operators. See \eqref{OO dT} for an explicit example. In order to get rid of the integrals in \eqref{tilde V0 2}, one integrates by part these differential operators such as to make them act on the correlators $\langle \epsilon\,...\, \epsilon \rangle$ instead. Equation \eqref{V0 equal V0} is derived provided that the net result of this integration by part and elimination of the integrals is
	\begin{align}
	\label{missing proof}
	\frac{1}{n!} \int \prod_{i=1}^n d^2z_i\, \langle VV \partialbar T_1\,...\,. \partialbar T_n \rangle_c\, \langle \epsilon_1\,...\, \epsilon_n\,...\rangle_{\tilde c}\, \langle \,...\, \rangle_c=(z_{12})^{-2h_V}\langle \mathcal{B}_{h_V}^{(n)}\,...\,\rangle_c\,.
	\end{align}
	The validity of this relation was shown explicitly for $n=1,2$ in \cite{Nguyen:2021jja}, however a systematic proof for generic $n$ is still missing at this time. Assuming \eqref{missing proof} does indeed hold, we can conclude that the formula for the Virasoro identity block \eqref{identity block} together with the accompanying prescriptions reviewed in the introduction are just another a way to write down the sum \eqref{Feynman sum} over all Feynman diagrams involving stress tensor exchanges between the external operators $V,W$. 
	
	Note that the very same reasoning also applies to \textit{star channel} contributions \cite{Anous:2020vtw} to correlators $\langle \prod_{i=1}^n O_i O_i \rangle$ of $n$ pairs of identical operators with ($O_i\neq O_j$ for $i\neq j$). Again a complete proof requires to demonstrate the generic validity of \eqref{missing proof} with $n$ bilocal insertions.
	
	This provides a formal understanding of the reparametrization formalism in terms of the more standard concepts of generating functionals and Feynman diagrams. Beyond putting it on firmer theoretical grounds, this also provides avenues for further developments. A first important point discussed at length in \cite{Nguyen:2021jja} is the treatment of UV divergences that appear when evaluating \eqref{identity block}. Having recast the reparametrization formalism in terms of Feynman diagrams, one should be able to apply standard renormalization techniques and translate them back to that formalism. In this regard it should be emphasized that in nonperturbative treatments of conformal field theory there are no UV divergences at all (see for instance \cite{Poland:2018epd}), while divergences arise in the approaches discussed in this paper precisely because they are set up in the framework of effective field theory. Another interesting avenue is a generalization of the formalism which would permit to compute Virasoro blocks associated with the exchange of a generic operator $O$ and its descendants. This is the subject of the last section.
	
	\section{Towards generic Virasoro blocks}
	\label{sec:generic blocks}
	In this paper I have provided a basis of understanding for the reparametrization formalism from first principles. In this section I briefly discuss an extension of the formalism which would allow to compute generic Virasoro blocks. I leave a detailed study to future investigation. 
	
	We are after the analogue of formula \eqref{identity block} for the exchange of an arbitrary operator $O$ and its descendants between the four external operators $O_1, O_2, O_3, O_4$. For concreteness we will focus our attention on the $s$-channel block $O_1 O_2 \rightarrow O+ desc. \rightarrow O_3 O_4$. 
	
\begin{figure}
		\centering
		\begin{subfigure}[b]{.49\textwidth}
			\centering
			\includegraphics[scale=0.4]{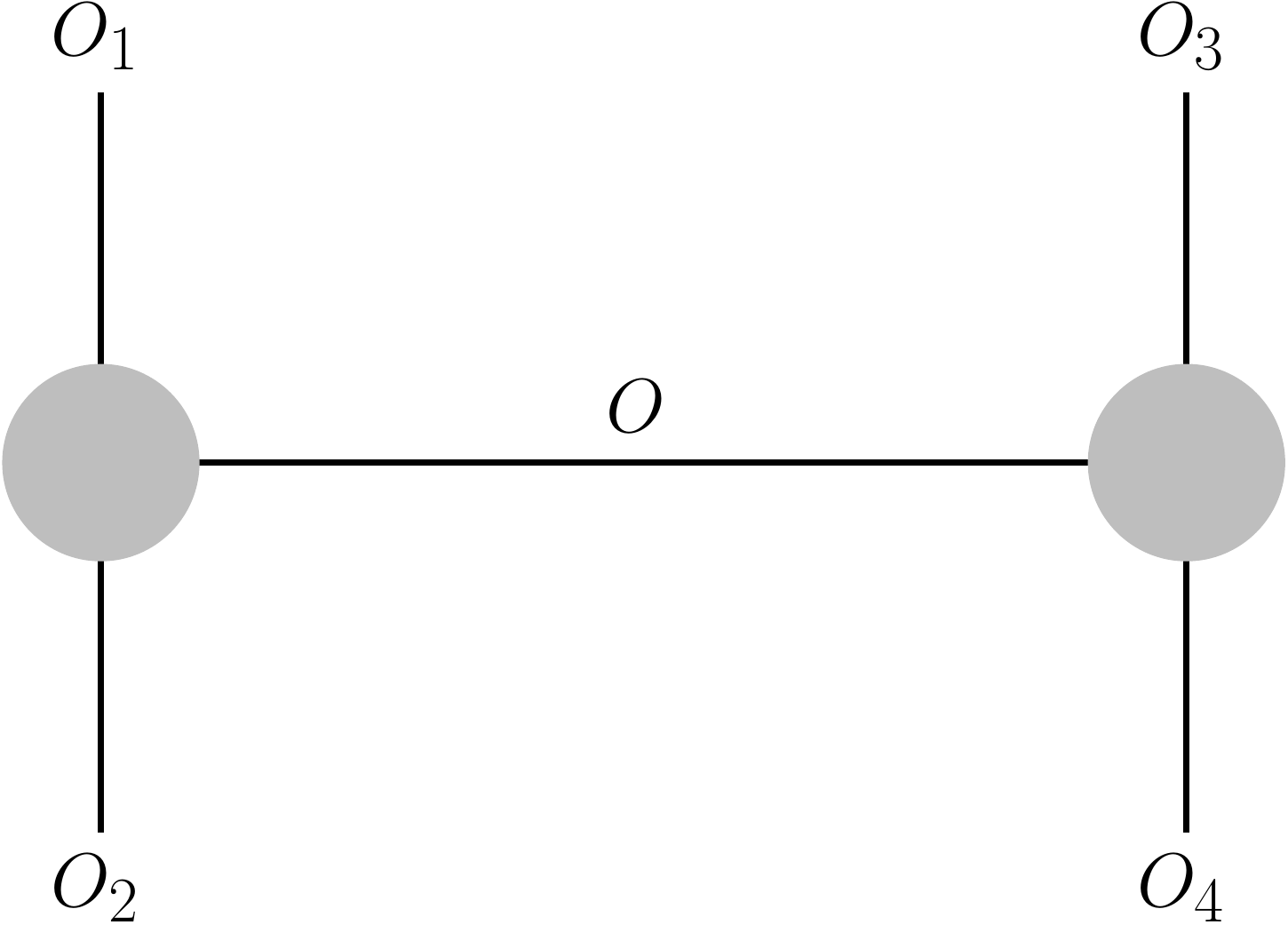}
		\end{subfigure}
		\begin{subfigure}[b]{.49\textwidth}
			\centering
			\includegraphics[scale=0.4]{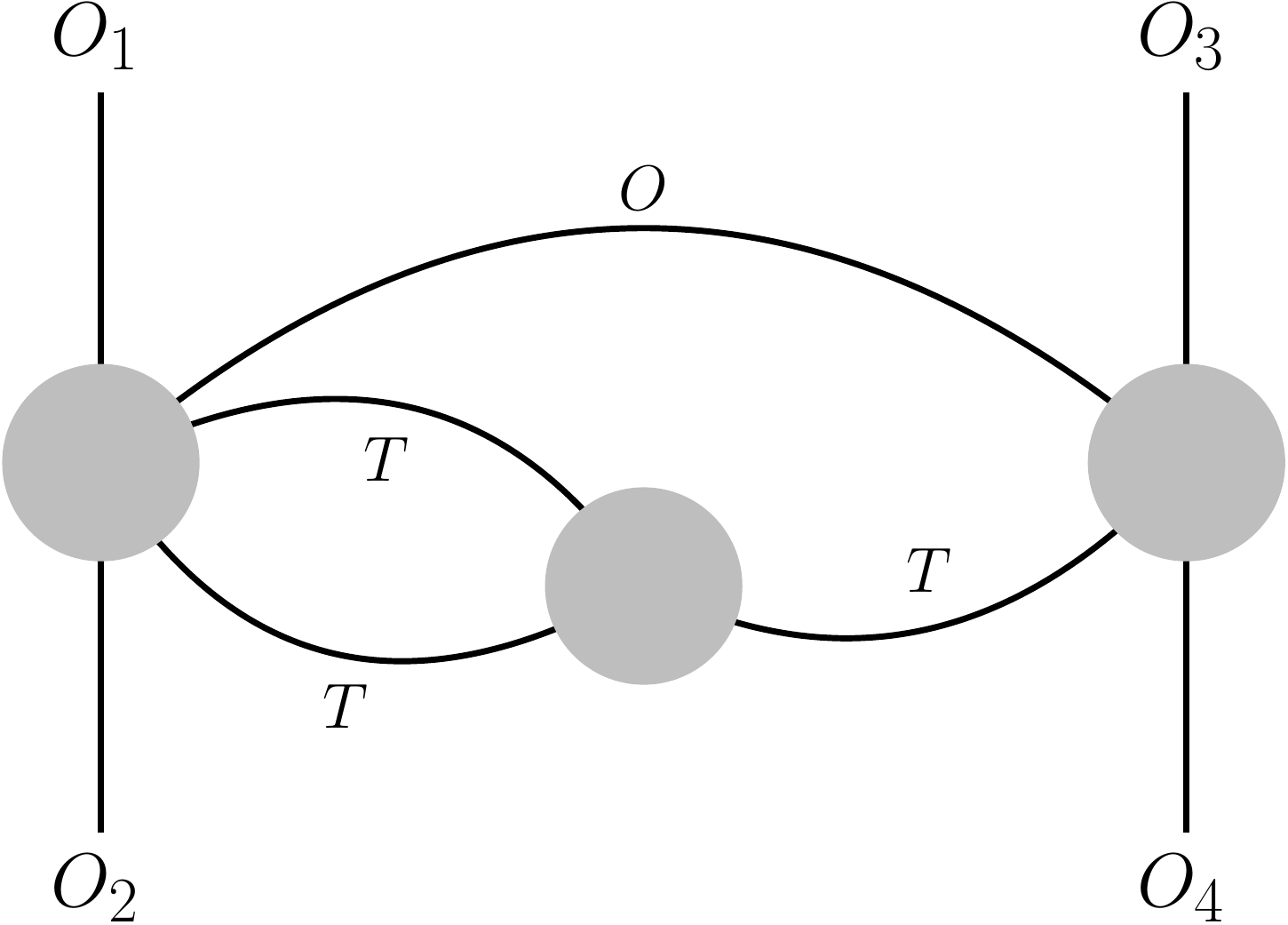}
		\end{subfigure}
		\caption{Feynman diagrams corresponding to the exchange of a primary operator $O$ and its descendants.}
		\label{fig:Feynman}
	\end{figure}

	We shall reverse the logic of the preceding section, i.e., we shall first write down the relevant sum of Feynman diagrams and turn it into an expression that is suitable to the reparametrization formalism. The relevant Feynman diagrams are those involving the exchange of one $O$ and an arbitrary number of $T$'s as in figure~\ref{fig:Feynman}. The relevant block contribution to the four point function $\langle O_1O_2O_3O_4 \rangle$ should therefore be given by
	\begin{equation}
	\label{VO}
	\mathcal{F}_O=\sum_{n,m}\frac{1}{n!m!} \int dM d^2y\, \langle O_1O_2 O(y) \hat T_1\,...\, \hat T_n\rangle_c\, \langle T_1\,...\, T_m \rangle_{\tilde c}\, \langle O_3O_4 \hat O(y) \hat T_1\,...\, \hat T_m\rangle_c\,.
	\end{equation}
	Again hatted operators correspond to amputated legs, according to
	\begin{equation}
	\langle\, ...\, \cO(z)\,...\, \rangle=\int d^2w\, \langle \cO(z)\cO(w)\rangle\, \langle\, ...\, \hat \cO(w)\,...\, \rangle\,,
	\end{equation}
	which is also the defining relation of the \textit{shadow transform} $\hat O(z)\equiv \text{Sh}_z\, O(z)$ \cite{Osborn:2012vt}. 
	Note that connecting two amputated legs via a propagator is equivalent to connecting an amputated leg to one which is not. I adopted the second option in writing \eqref{VO}. 
	
	Just as in the previous section, we can now use \eqref{hat T}, \eqref{eq:epsilon propagators} and integration by part in order to rewrite \eqref{VO} as
	\begin{equation}
	\mathcal{F}_O=\sum_{n,m}\frac{1}{n!m!} \int dM d^2y\, \langle O_1O_2 O(y) \partialbar T_1\,...\, \partialbar T_n\rangle_c\, \langle \epsilon_1\,...\, \epsilon_m \rangle_{\tilde c}\, \langle O_3O_4 \hat O(y) \partialbar T_1\,...\, \partialbar T_m\rangle_c\,.  
	\end{equation}
	The correlators $\langle O_1O_2 O \partialbar T\,...\, \partialbar T\rangle_c$ are the functional derivatives of the generating functional $Z_{3,c}[\mu]$ given in \eqref{Psi3}. Therefore, and provided that an analogue of \eqref{missing proof} applies here as well, we can write the Virasoro block of interest as
	\begin{equation}
	\label{new formula}
	\mathcal{F}_O=\int d^2y\, \langle Z_{3,c}(1,2,y)\, \text{Sh}_y\, Z_{3,c}(y,3,4) \rangle_c\,.
	\end{equation}
	This formula is significantly more involved than its identity block counterpart \eqref{identity block}. Beyond the replacement of $Z_{2,c}$ by $Z_{3,c}$, it features a shadow transform implemented by $\text{Sh}_y$ together with an explicit integral accounting for the propagation of the primary operator $O$. The practicality of this formula might therefore be limited. A first natural test of the formula \eqref{new formula} would be to recover the leading term in the $1/c$ expansion of $\mathcal{F}_O$, namely the global conformal block associated with the exchange of $O$ \cite{Dolan:2000ut,Dolan:2003hv}.
	
	\section*{Acknowledgments}
	I thank Felix Haehl, Mark Mezei and Gideon Vos for stimulating discussions. This work is supported by the STFC grants ST/P000258/1 and ST/T000759/1.
	
	\appendix
	
	\section{Diffeomorphism Ward identities}
	\label{appendix}
	Following the approach described in \cite{Yoshida:1988xm}, I provide a derivation of the diffeomorphism Ward identities satisfied by the generating functionals for correlation functions of the form
	\begin{equation}
	\label{corelators}
	\langle \Om\, T(w_1)...T(w_n) \rangle\,,
	\end{equation}
	where $\Om$ is defined in \eqref{O string} to be a string of $m$ primary operators. For a given $\Om$, the associated chiral generating functional is given by
	\begin{equation}
	Z\left[\mu \right] = \sum_{n=0}^\infty \frac{(-\pi)^{-n}}{n!} \int d^2w_1...\,d^2w_n\, \mu(w_1,\wbar_1)...\mu(w_n,\wbar_n) \langle T(w_1)...T(w_n) \Om  \rangle\,,
	\end{equation}
	such that correlators \eqref{corelators} can be obtained by functional differentiation of $Z[\mu]$ with respect to the background field $\mu$. Importantly they also satisfy the conformal Ward identities \cite{Belavin:1984vu}
	\begin{gather}
	\label{eq:conf Ward}
	\begin{split}
	&\langle T(z)T(w_1)...T(w_n) \Om\rangle=\sum_{i=1}^n \frac{c}{2(z-w_i)^4}\, \langle  T(w_1)...\cancel{T(w_i)}...T(w_n) \Om \rangle\\
	&+\left[\sum_{i=1}^n  \left(\frac{2}{(z-w_i)^2}+\frac{\partial_{w_i}}{z-w_i}\right)+\sum_{j=1}^m \left(\frac{h_j}{(z-z_j)^2}+\frac{\partial_{z_j}}{z-z_j}\right)\right] \langle T(w_1)...T(w_n) \Om \rangle\,,
	\end{split}
	\end{gather}
	which, thanks to the magic identity \eqref{eq:magic}, directly yields the diffeomorphism Ward identity
	\begin{gather}
	\label{eq:diff Ward}
	\frac{1}{\pi}\langle \partialbar T(z)T_1\,...\,T_n \Om \rangle=-\sum_{i=1}^n \frac{c}{12}\, \partial^3 \delta(z-w_i) \, \langle T_1\,...\,\cancel{T_i}\,...\,T_n \Om \rangle\\
	\nonumber
	-\left[\sum_{i=1}^n \left(2 \partial \delta(z-w_i)-\delta(z-w_i)\partial_{w_i}\right)+\sum_{j=1}^m \left(h_j \partial\delta(z-z_j)-\delta(z-z_j)\partial_{z_j}\right)\right] \langle T_1\,...\,T_n \Om \rangle\,.
	\end{gather}
	This in turn translates into a Ward identity for the the generating functional,
	\begin{equation}
	\left(\partialbar- \mu \partial-2 \partial \mu \right) \frac{\delta Z\left[\mu\right]}{\delta \mu(z,\zbar)}=\left[-\frac{c}{12\pi} \partial^3 \mu+\sum_{j=1}^m \left(h_j \partial\delta(z-z_j)-\delta(z-z_j)\partial_{z_j}\right) \right]Z\left[\mu\right]\,.
	\end{equation}
	We can alternatively write it in terms of the quasi-conformal mapping $f$,
	\begin{equation}
	\label{eq:Ward identity}
	\partial f(z,\zbar)\, \frac{\delta Z\left[\mu\right]}{\delta f(z,\zbar)}=\left[\frac{c}{12\pi} \partial^3 \mu-\sum_{j=1}^m \left(h_j \partial\delta(z-z_j)-\delta(z-z_j)\partial_{z_j}\right) \right]Z\left[\mu\right]\,,
	\end{equation}
	since the functional chain rule yields
	\begin{equation}
	\begin{split}
	\frac{\delta}{\delta f(z,\zbar)}&=\int d^2w\, \frac{\delta \mu(w,\wbar)}{\delta f(z,\zbar)} \frac{\delta}{\delta \mu(w,\wbar)}\\
	&=\int d^2w \left(\frac{1}{\partial f}\, \partial_{\wbar} \delta(w-z)-\frac{\partialbar f}{(\partial f)^2}\, \partial_w \delta(w-z) \right) \frac{\delta}{\delta \mu(w,\wbar)}\\
	&=-\frac{1}{\partial f}\left(\partialbar -\mu \partial -2\partial \mu\right) \frac{\delta}{\delta \mu(z,\zbar)}\,.
	\end{split}
	\end{equation}
	I will now discuss the solutions to the Ward identity \eqref{eq:Ward identity}.
	
	\paragraph{No operator.}
	Without any operator insertion of reference, $\Om=\emptyset$, the explicit solution for the generating functional is given by 
	\begin{equation}
	Z_0\left[\mu\right]=\exp\left[-\frac{c}{24\pi} \int d^2z\,  \frac{\partialbar f}{\partial f}\, \partial^2 \ln \partial f\right]\,,
	\end{equation}
	which allows to compute stress tensor correlators $\langle T\,...\, T \rangle$.\\
	
	We can use $Z_0[\mu]$ to simplify the determination of $Z[\mu]$ associated with a generic string of operators $\Om$. Indeed writing
	\begin{equation}
	Z[\mu]=Z_0[\mu]\, Z_c[\mu]\,,
	\end{equation}
	the Ward identity \eqref{eq:Ward identity} reduces to
	\begin{equation}
	\label{connected Ward identity}
	\partial f(z,\zbar)\, \frac{\delta Z_c[\mu]}{\delta f(z,\zbar)}
	=-\sum_{j=1}^m \left(h_j \partial\delta(z-z_j)-\delta(z-z_j)\partial_{z_j}\right) Z_c[\mu]\,.
	\end{equation}
	The natural interpretation is that $Z_c[\mu]$ generates \textit{connected} correlations between $\Om$ and the stress tensors insertions.
	
	\paragraph{One operator.} The correlators \eqref{corelators} are all trivial since $\langle \cO \rangle=0$ as a result of conformal invariance of the vacuum. 
	
	\paragraph{Two operators.} For a pair of identicl operators $\Om=\cO(w_1) \cO(w_2)$ with conformal weight $h$, the solution to the Ward identity \eqref{connected Ward identity} is given by the reparametrized two-point function
	\begin{equation}
	Z_{2,c}[\mu]=\left(\frac{\partial f(z_1,\zbar_1) \partial f(z_2,\zbar_2)}{(f(z_1,\zbar_1)-f(z_2,\zbar_2))^2} \right)^h\equiv \left(\frac{\partial f_1\, \partial f_2}{(f_1-f_2)^2} \right)^h\,.
	\end{equation}
	Indeed, we have
	\begin{equation}
	\begin{split}
	\frac{\delta Z_{2,c}[\mu]}{\delta f(z,\zbar)} &=\left[ \frac{h\,  \partial_{z_1} \delta(z-z_1)}{\partial f_1}- \frac{2h\, \delta(z-z_1)}{f_1-f_2}+ (z_1 \leftrightarrow z_2) \right]Z_{2,c}[\mu]\\
	&=\left[\frac{h\, \partial_{z_1} \delta(z-z_1)}{\partial f_1}-h\delta(z-z_1) \frac{\partial^2f_1}{(\partial f_1)^2}+\frac{\delta(z-z_1)}{\partial f_1}\partial_{z_1}+ (z_1 \leftrightarrow z_2) \right]Z_{2,c}[\mu]\\
	&=-\frac{1}{\partial f(z,\zbar)}\left[h\, \partial_z \delta(z-z_1)-\delta(z-z_1)\partial_{z_1}+ (z_1 \leftrightarrow z_2) \right]Z_{2,c}[\mu]\,,
	\end{split}
	\end{equation}
	where in the first equality we used
	\begin{equation}
	\partial_{z_1}Z_{2,c}[\mu]=h\left(\frac{\partial^2f_1}{\partial f_1}- \frac{2\partial f_1}{f_1-f_2}\right)Z_{2,c}[\mu]\,,
	\end{equation}
	while in the second equality we used
	\begin{equation}
	\frac{\partial_{z_1} \delta(z-z_1)}{\partial f_1}-\delta(z-z_1) \frac{\partial^2f_1}{(\partial f_1)^2}=\partial_{z_1}\left(\frac{\delta(z-z_1)}{\partial f_1}\right)=\frac{\partial_{z_1} \delta(z-z_1)}{\partial f(z,\zbar)}=-\frac{\partial_z \delta(z-z_1)}{\partial f(z,\zbar)}\,.
	\end{equation}
	
	\paragraph{Three operators.} For $\Om=\cO_1(w_1)\cO_2(w_2)\cO_3(w_3)$, the solution to the Ward identity \eqref{connected Ward identity} is the reparametrized 3-point function
	\begin{equation}
	Z_{3,c}[\mu]=\frac{(\partial f_1)^{h_1} (\partial f_2)^{h_2} (\partial f_3)^{h_3}}{(f_1-f_2)^{h_1+h_2-h_3}(f_1-f_3)^{h_1+h_3-h_2}(f_2-f_3)^{h_2+h_3-h_1}}\,.
	\end{equation}
	The demonstration is similar to that for $Z_{2,c}[\mu]$ above and is left as an exercise to the reader.
	
	\paragraph{More operators.} One might conjecture that the generating functional associated with a longer string of operators $\Om$ is the reparametrized correlation function $\langle \Om \rangle$. Although it would be interesting to prove or disprove it, I will not investigate this further here as they are not needed in this work.
	
	\bibliography{bibl}
	\bibliographystyle{JHEP}
\end{document}